\documentclass[a4paper,11pt]{article}
\usepackage{pos}

\title{The SkyLLH framework for IceCube point-source search}
 \ShortTitle{The SkyLLH framework for IceCube point-source search}

\author{The IceCube Collaboration \\{\normalsize \normalfont(a complete list of authors can be found at the end of the proceedings)}}

\emailAdd{tomas.kontrimas@icecube.wisc.edu}

\abstract{
Hypothesis tests based on unbinned log-likelihood (LLH) functions are a common technique used in multi-messenger astronomy, including IceCube's neutrino point-source searches. We present the general Python-based tool "SkyLLH", which provides a modular framework for implementing and executing log-likelihood functions to perform data analyses with multi-messenger astronomy data. Specific SkyLLH framework features for a new and improved time-integrated IceCube point-source analysis are highlighted, including the support for kernel density estimation (KDE) based probability density functions. In addition, the support for a variety of point-source analysis types, such as stacked and time-variable searches, will be presented.

\vspace{4mm}
{\bfseries Corresponding authors:}
Tomas Kontrimas$^{1*}$, Martin Wolf$^{1}$\\
{$^{1}$ \itshape Physik-department, Technische Universität München, D-85748 Garching, Germany}\\[4mm]
$^*$ Presenter

\FullConference{37$^{\rm{th}}$ International Cosmic Ray Conference (ICRC 2021)\\
		July 12th -- 23rd, 2021\\
		Online -- Berlin, Germany}

}

\begin{document}
\maketitle

\section{Introduction}

The unbinned likelihood formalism is used widely in multi-messenger and neutrino astronomy as a statistical test \cite{braunMethodsPointSource2008}. For testing two complementary hypotheses, the likelihood ratio test is expressed as a ratio of likelihoods given each hypothesis parameters $\vec{\theta}$ from an entire parameter space $\Theta$
\begin{equation}
    \lambda ( \vec{D} ) = \frac{\sup_{\vec{\theta} \in \Theta_0} \mathcal{L} ( \vec{\theta} | \vec{D} )}{\sup_{\vec{\theta} \in \Theta} \mathcal{L}( \vec{\theta} | \vec{D} )},
    \label{eq:likelihood_ratio}
\end{equation}
where hypotheses $H_0:\vec{\theta} \in \Theta_0$ and $H_1: \vec{\theta} \in \Theta_0^c$ are called null and alternative hypotheses, respectively. The resulting $\lambda (\vec{D})$ value indicates whether the given data is compatible with the null hypothesis or not.

The unbinned likelihood function is generally expressed as $\mathcal{L}=\prod_i p(\vec{x}_i)$, with assigned probability $p(\vec{x}_i)$ to each observed event $\vec{x}_i$, where the total likelihood value is given by a product. In IceCube neutrino point-source searches, the probability of observed event is written as a two-component likelihood function
\begin{equation}
    \mathcal{L} \left( n_s, \vec{p}_s | D \right) = \prod_{i=1}^N \left[ \frac{n_s}{N} S_i \left( \vec{p}_s \right) + \left( 1 - \frac{n_s}{N} \right) B_i \right],
    \label{eq:likelihood}
\end{equation}
where $n_s$ is the number of signal events in the data sample $D$ of $N$ total events. The set of source model parameters is denoted as $\vec{p}_s$ and usually contains source position $\vec{d}_{\text{src}} = (\alpha_{\text{src}}, \delta_{\text{src}})$ and spectral index $\gamma$. $S_i$ and $B_i$ are the value of the signal and background probability density function (PDF) for the $i$th data event, respectively. Because the number of signal events in the data sample is usually small, the background and signal PDFs can be built either from scrambled experimental or Monte-Carlo simulation data.

The log likelihood ratio test statistic is expressed as
\begin{equation}
    \mathcal{TS} = -2 \log \lambda \left( n_s, \vec{p}_s | D \right) = -2 \sup_{n_s, \vec{p}_s} \left\{ \log \left[ \frac{\mathcal{L} \left( n_s = 0 | D \right)}{\mathcal{L} \left( n_s, \vec{p}_s | D \right)} \right] \right\},
\end{equation}
where the null hypothesis is defined as an observation of zero signal events and the alternative hypothesis --- observation of $n_s > 0$. The logarithm is used outside of the likelihood ratio for the numerical stability. In the limit of high statistics, the Wilk's approximation \cite{wilksLargeSampleDistributionLikelihood1938} can be used to approximate it. The resulting $\mathcal{TS}$ distribution follows a $k$ number of degree of freedom $\chi_k^2$-distribution when hypotheses parameters are sufficiently away from boundaries.

\section{The SkyLLH Framework}

The SkyLLH framework is an open-source Python3-based tool licensed under the GPLv3 license\footnote{https://www.gnu.org/licenses/gpl-3.0.txt}.
It is available on the open-source IceCube Neutrino Observatory GitHub repository\footnote{https://github.com/icecube/skyllh} and provides a modular framework for implementing custom likelihood functions and executing log-likelihood ratio hypothesis tests. Because the framework is designed to be detector independent, it is easy to extend it and perform analyses with combined multi-messenger data from IceCube, Fermi, ANTARES, KM3NeT, etc. observatories.

Such framework modularity was achieved by defining common interfaces for mathematical components of the log-likelihood function utilizing object-oriented-programming (OOP) techniques. The classes structure is tied to the mathematical objects of the likelihood (ratio) function, for example, an $\texttt{Analysis}$ object can be naturally built from PDFs $\rightarrow$ PDFRatios $\rightarrow$ LLHRatio objects. This allows to define a specific analysis by choosing from already provided components or extending them with custom properties, if needed. The specific complementary classes for the IceCube Neutrino Observatory detector are provided in the $\texttt{i3}$ module and a private IceCube github repository\footnote{https://github.com/icecube/i3skyllh} containing pre-defined common analyses and datasets.

The main framework code is split into four modules. The $\texttt{core}$ module holds classes defining the detector independent mathematical log-likelihood framework by utilizing Python's abstract base class (ABC) module. As mentioned earlier, IceCube specific classes derived from the $\texttt{core}$ module are in the $\texttt{i3}$ module. They extend signal and background events generators, provide detector signal yield calculation, models of PDFs and PDFRatios, and common coordinate transformations. The $\texttt{physics}$ module contains definitions of source hypothesis and its flux models. The $\texttt{plotting}$ module contains utility functions for plotting the generated analysis objects (PDFRatios and PDFs) and the calculated trial data.

The analysis definition process in SkyLLH is realized by creating an $\texttt{Analysis}$ object with desired properties. For the users' convenience, the $\texttt{Analysis}$ object can be simply created by running a function from a list of already predefined $\texttt{create\_analysis}$ functions of commonly used analysis types, or they can serve as a base for a custom analysis definition. The important $\texttt{create\_analysis}$ function parameters are $\texttt{datasets}$, $\texttt{minimizer\_impl}$, $\texttt{source}$, reference power-law flux model normalization, energy, and spectral index.

The $\texttt{datasets}$ parameter contains a list of $\texttt{Dataset}$ instances. SkyLLH has pre-defined functions for creating $\texttt{DatasetCollection}$ of datasets, which are currently used in most of the ongoing IceCube analyses. The desired dataset collection, containing a list of the same version datasets, can simply be loaded from the provided $\texttt{data\_samples}$ dictionary. Predefined datasets can be easily customized by adding additional experimental, Monte-Carlo, and (or) auxiliary data definitions. In addition to the data definitions, the $\texttt{Dataset}$ object can contain data preparation functions, which modify data after it is loaded, and data field renaming dictionaries for experimental and Monte-Carlo data to standardize the naming scheme between different dataset collections.

The likelihood developer can either define a custom minimizer implementation or choose it from provided Scipy\footnote{https://docs.scipy.org/doc/scipy/reference/generated/scipy.optimize.minimize.html}, Newton-Raphson, and iminuit\footnote{https://iminuit.readthedocs.io/en/stable/reference.html} implementations.
The source is defined by e.g. a $\texttt{PointLikeSource}$ instance at a given location in the sky. If the likelihood developer wants to change the source position after the $\texttt{Analysis}$ object is created, the $\texttt{Analysis}$ class has a $\texttt{change\_source}$ method, which applies necessary changes to all the source dependent objects of the $\texttt{Analysis}$. For the source flux model the $\texttt{PowerLawFlux}$ is created with given flux normalization, energy, and spectral index values. 

\section{Kernel Density Estimator for A New Search for Neutrino Point Sources with IceCube analysis}

Using the SkyLLH framework a new point-source analysis with the IceCube data was developed. IceCube is a cubic-kilometer neutrino detector installed in the ice at the geographic South Pole \cite{aartsenIceCubeNeutrinoObservatory2017} between depths of 1450 m and 2450 m, completed in 2010. Reconstruction of the direction, energy and flavor of the neutrinos relies on the optical detection of Cherenkov radiation emitted by charged particles produced in the interactions of neutrinos in the surrounding ice or the nearby bedrock. The new search for neutrino point-sources improves the accuracy of the statistical analysis, especially in the low energy regime \cite{new_ps:2021icrc}, in comparison to previous analyses. It has a new likelihood description based on kernel density estimation (KDE) method, which allows to replace analytical expressions by non-parametrically inferring probability density functions.

The multi-dimensional KDE based PDFs are defined as
\begin{equation}
    P_{\text{KDE}}(\vec{x}) = \frac{1}{N h} \sum_{i=1}^N K \left( \frac{\vec{x} - \vec{x}_i}{h} \right),
\end{equation}
where $h$ is a smoothing parameter called bandwidth, $K(x)$ is a kernel function and $\vec{x}_i$ is $i$th event in a data sample.
The new point-source analysis replaces signal and background PDF terms in the likelihood expression (equation \ref{eq:likelihood}) with
\begin{equation}
    \label{eq:final_signal}
    f_S\left(\hat{E}_{\mu},\,\hat{\vec{d}},\,\hat{\sigma}\right) \approx \frac{1}{2\pi \, \sin\psi}\,f_S\left(\hat{\psi}\,|\,\hat{\sigma},\,\hat{E}_{\mu},\,\gamma \right) \cdot f_S\left(\hat{E}_{\mu}\,|\,\delta_{src},\,\gamma\right),   
\end{equation}

\begin{equation}
\label{eq:llh_bkg}
    f_B(\hat{\vec{d}}_i, \hat{\sigma}_i, \hat{E}_{\mu,\,i}) = \frac{1}{2\pi} f_B(\sin\hat{\delta}_i, \hat{\sigma}_i, \hat{E}_{\mu,\,i}),
\end{equation}
respectively. It uses three observables: the estimated muon energy $\hat{E}_{\mu}$, the reconstructed muon direction $\hat{\vec{d}}$ and its estimated uncertainty $\hat{\sigma}$, and angular distance $\hat{\psi}=||\hat{\vec{d}}_i - \vec{d}_{src}||$ between reconstructed direction and source position. The new likelihood no longer relies on the analytical Gaussian approximation of signal PDF's spatial term, which was used in previous IceCube analyses \cite{braunMethodsPointSource2008}.

This likelihood construction requires generation of KDE based signal PDFs on a power-law flux index $\gamma$ grid.
They are generated from the Monte-Carlo data using an internally developed KDE\_tool based on Meerkat \cite{poluektovKernelDensityEstimation2015} package, which uses cross validation technique in order to find an optimal KDE bandwidth $h$ on a coarse $\gamma$ grid. After the optimal bandwidths are found, KDEs are generated on a fine $\gamma$ grid using interpolated optimal bandwith values. Because the KDE evaluation is slow, they are evaluated only once on a very fine parameter grid, the result is interpolated and saved as penalized B-splines using the photospline\footnote{https://github.com/icecube/photospline} package.

In order to use the new likelihood function in SkyLLH, the framework was extended to provide linear and parabola grid manifold interpolation methods. They selectively load and interpolate penalized B-splines in-between $\gamma$ grid values while minimizing the log-likelihood function. The photosplines are defined as auxiliary $\texttt{Dataset}$ object properties.

\section{Stacking}


In general, a likelihood ratio test can be done not only for a single point-source, but also for a set of $K$ stacked sources in a weighted fashion. This can be viewed as an extension to the single point-source search method. The sources must be weighted according to their signal detection efficiency $Y_{s, k}$, and a relative strength weight of the $k$th source $W_k$, with $\sum_{k=1}^K W_k = 1$. The $W_k$ is a theoretical weight independent of position and spectral index of the source that accounts for different properties of the individual sources. The combined signal PDF is then given as
\begin{equation}
    S_i(\vec{p}_s) \equiv \frac{\sum_{k=1}^K W_k Y_s(\vec{x}_{s_k}, \vec{p}_{s_k}) S_i(\vec{p}_{s_k})}{\sum_{k=1}^K W_k Y_s (\vec{x}_{s_k}, \vec{p}_{s_k})}.
\end{equation}

In SkyLLH the source hypothesis group class provides a data container to describe a group of sources that share the same flux model, detector signal yield, and signal generation implementation methods. It also supports a definition of relatively weighted multiple sources. The stacking analysis can be constructed by setting up a \texttt{TimeIntegratedMultiDatasetMultiSourceAnalysis} object.

\section{Summary and Outlook}

The SkyLLH framework is being developed within the IceCube collaboration as a standard tool to search for neutrino emitting sources in the Universe. The implementation of generalized concepts in terms of source hypothesis and hypothesis parameter definition makes it easy to use the SkyLLH framework also for searches of other messenger particles in other experiments. Whenever a likelihood ratio test as given in equation \ref{eq:likelihood_ratio} with celestial data has to be performed, SkyLLH is a suitable tool. Possible future applications of SkyLLH could be combined analyses of same-kind messenger particle data, for instance from different neutrino telescopes like IceCube and ANTARES / KM3NeT, or of different messenger particle data of neutrinos and gamma-rays, for instance from IceCube and Fermi/LAT.





\bibliographystyle{ICRC}
\bibliography{references}

\providecommand{\href}[2]{#2}\begingroup\raggedright\begin{thebibliography}{1}

\bibitem{braunMethodsPointSource2008}
J.~Braun, J.~Dumm, F.~De~Palma, C.~Finley, A.~Karle, and T.~Montaruli
  \href{http://dx.doi.org/10.1016/j.astropartphys.2008.02.007}{{\em Astropart.
  Phys.} {\bfseries 29} (2008) 299--305}.

\bibitem{wilksLargeSampleDistributionLikelihood1938}
S.~S. Wilks \href{http://dx.doi.org/10.1214/aoms/1177732360}{{\em Annals Math.
  Statist.} {\bfseries 9} no.~1, (1938) 60--62}.

\bibitem{aartsenIceCubeNeutrinoObservatory2017}
{\bfseries IceCube} Collaboration, M.~G. Aartsen {\em et~al.}
  \href{http://dx.doi.org/10.1088/1748-0221/12/03/P03012}{{\em JINST}
  {\bfseries 12} no.~03, (2017) P03012}.

\bibitem{new_ps:2021icrc}
{\bfseries IceCube} Collaboration {\em PoS} {\bfseries ICRC2021} (these
  proceedings) 1138.

\bibitem{poluektovKernelDensityEstimation2015}
A.~Poluektov \href{http://dx.doi.org/10.1088/1748-0221/10/02/P02011}{{\em
  JINST} {\bfseries 10} (2015) P02011}.

\end{thebibliography}\endgroup



\clearpage
\section*{Full Author List: IceCube Collaboration}




\scriptsize
\noindent
R. Abbasi$^{17}$,
M. Ackermann$^{59}$,
J. Adams$^{18}$,
J. A. Aguilar$^{12}$,
M. Ahlers$^{22}$,
M. Ahrens$^{50}$,
C. Alispach$^{28}$,
A. A. Alves Jr.$^{31}$,
N. M. Amin$^{42}$,
R. An$^{14}$
% ...
\\

\noindent
% ...
$^{12}$ Universit{\'e} Libre de Bruxelles, Science Faculty CP230, B-1050 Brussels, Belgium \\
% ...
$^{14}$ Department of Physics and Laboratory for Particle Physics and Cosmology, Harvard University, Cambridge, MA 02138, USA \\
% ...
$^{17}$ Department of Physics, Loyola University Chicago, Chicago, IL 60660, USA \\
$^{18}$ Dept. of Physics and Astronomy, University of Canterbury, Private Bag 4800, Christchurch, New Zealand \\
% ...
$^{22}$ Niels Bohr Institute, University of Copenhagen, DK-2100 Copenhagen, Denmark \\
% ...
$^{28}$ D{\'e}partement de physique nucl{\'e}aire et corpusculaire, Universit{\'e} de Gen{\`e}ve, CH-1211 Gen{\`e}ve, Switzerland \\
% ...
$^{31}$ Karlsruhe Institute of Technology, Institute for Astroparticle Physics, D-76021 Karlsruhe, Germany  \\
% ...
$^{42}$ Bartol Research Institute and Dept. of Physics and Astronomy, University of Delaware, Newark, DE 19716, USA \\
% ...
$^{50}$ Oskar Klein Centre and Dept. of Physics, Stockholm University, SE-10691 Stockholm, Sweden \\
% ...
$^{59}$ DESY, D-15738 Zeuthen, Germany \\
% ...

\subsection*{Acknowledgements}

\noindent
USA {\textendash} U.S. National Science Foundation-Office of Polar Programs,
U.S. National Science Foundation-Physics Division,
U.S. National Science Foundation-EPSCoR,
Wisconsin Alumni Research Foundation,
Center for High Throughput Computing (CHTC) at the University of Wisconsin{\textendash}Madison,
Open Science Grid (OSG),
Extreme Science and Engineering Discovery Environment (XSEDE),
Frontera computing project at the Texas Advanced Computing Center,
U.S. Department of Energy-National Energy Research Scientific Computing Center,
Particle astrophysics research computing center at the University of Maryland,
Institute for Cyber-Enabled Research at Michigan State University,
and Astroparticle physics computational facility at Marquette University;
Belgium {\textendash} Funds for Scientific Research (FRS-FNRS and FWO),
FWO Odysseus and Big Science programmes,
and Belgian Federal Science Policy Office (Belspo);
Germany {\textendash} Bundesministerium f{\"u}r Bildung und Forschung (BMBF),
Deutsche Forschungsgemeinschaft (DFG),
Helmholtz Alliance for Astroparticle Physics (HAP),
Initiative and Networking Fund of the Helmholtz Association,
Deutsches Elektronen Synchrotron (DESY),
and High Performance Computing cluster of the RWTH Aachen;
Sweden {\textendash} Swedish Research Council,
Swedish Polar Research Secretariat,
Swedish National Infrastructure for Computing (SNIC),
and Knut and Alice Wallenberg Foundation;
Australia {\textendash} Australian Research Council;
Canada {\textendash} Natural Sciences and Engineering Research Council of Canada,
Calcul Qu{\'e}bec, Compute Ontario, Canada Foundation for Innovation, WestGrid, and Compute Canada;
Denmark {\textendash} Villum Fonden and Carlsberg Foundation;
New Zealand {\textendash} Marsden Fund;
Japan {\textendash} Japan Society for Promotion of Science (JSPS)
and Institute for Global Prominent Research (IGPR) of Chiba University;
Korea {\textendash} National Research Foundation of Korea (NRF);
Switzerland {\textendash} Swiss National Science Foundation (SNSF);
United Kingdom {\textendash} Department of Physics, University of Oxford.

\end{document}